\documentclass[aip,chaos,reprint,twocolumn,superscriptaddress,longbibliography,showpacs,floatfix]{revtex4-2}

\usepackage{graphicx}
\usepackage{amsmath}
\usepackage{amssymb}
\usepackage{epsf}
\usepackage{color}
\usepackage{mathtools}

\begin{document}

\title{Channel assisted noise propagation in a two-step cascade}

\author{Mintu Nandi}
\email{mintu.rs2022@chem.iiests.ac.in}
\affiliation{Department of Chemistry, Indian Institute of Engineering Science and Technology, Shibpur, Howrah 711103, India}

\author{Sudip Chattopadhyay}
\email{sudip@chem.iiests.ac.in}
\affiliation{Department of Chemistry, Indian Institute of Engineering Science and Technology, Shibpur, Howrah 711103, India}

\author{Somshubhro Bandyopadhyay}
\email{som@jcbose.ac.in}
\affiliation{Department of Physical Sciences, Bose Institute, EN 80, Sector V, Bidhan Nagar, Kolkata 700091, India}

\author{Suman K Banik}
\email{skbanik@jcbose.ac.in}
\affiliation{Department of Chemical Sciences, Bose Institute, EN 80, Sector V, Bidhan Nagar, Kolkata 700091, India}

\begin{abstract}
Signal propagation in biochemical networks is characterized by the inherent randomness in gene expression and fluctuations of the environmental components, commonly known as intrinsic and extrinsic noise, respectively. We present a theoretical framework for noise propagation in a generic two-step cascade (S$\rightarrow$X$\rightarrow$Y) regarding intrinsic and extrinsic noise.  We identify different channels of noise transmission that regulate the individual and the overall noise properties of each component. Our analysis shows that the intrinsic noise of S alleviates the general noise and information transmission capacity along the cascade. On the other hand, the intrinsic noise of X and Y acts as a bottleneck of information transmission. We also show a hierarchical relationship among the intrinsic noise levels of S, X, and Y, with S exhibiting the highest level of intrinsic noise, followed by X and then Y. This hierarchy is preserved within the two-step cascade, facilitating the highest information transmission from S to Y via X.
\end{abstract}

\date{\today}

\maketitle


\textbf{Noise and information transmission play decisive roles in elucidating the performance of biochemical networks in a fluctuating cellular environment. While noise can sometimes benefit cellular processes, it can also be detrimental. Decomposing noise into intrinsic and extrinsic components aids in understanding their impact on network performance. Information transmission, however, measures how well a network can adapt to a changing environment by conveying signals from one end of the network to another. Understanding how noise and information transmission interact is crucial for deciphering network behavior under noisy conditions. In the present work, we correlate the cellular noise with information transmission along a cascaded network. Our study highlights the hierarchical relationship among intrinsic noise levels in various components of the network, revealing their crucial role in optimizing information transmission. Furthermore, we examine how the interplay between intrinsic and extrinsic noise shapes the system's ability to adapt and maintain robustness in cellular response, providing insights into fundamental processes of biological signal transduction.}


\section{Introduction}
\label{s1}

Noise plays an essential role in shaping the dynamics of various cellular processes, including gene expression, transcriptional regulation, signal transduction pathways, and developmental pathways \cite{Kaern2005,Eldar2010,Tsimring2014}. These cellular processes adopt diverse strategies in response to the inevitable influence of noise. Comprehension of the precise impact of noise on cellular processes is a complex endeavor and needs individual attention. In this context, exploring the influence of noise on regulatory cascades could reveal crucial insights into the role of noise in cellular functioning. Regulatory cascades are a common occurrence in biological systems, manifesting in various forms, one of which is transcriptional cascades in \textit{Escherichia coli} and \textit{Saccharomyces cerevisiae} \cite{Shen-Orr2002,Rosenfeld2003}. Moreover, cascades are pivotal in directing the temporal sequencing of gene expression, contributing to critical processes like sporulation \cite{McAdams2003} and flagella formation \cite{Kalir2001}. In more complex organisms like Drosophila and sea urchins, the developmental programs heavily rely on precisely orchestrated cascaded processes to achieve intricate temporal coordination of events \cite{Arnone1997,Davidson2002}.

\begin{figure*}[t!]
\includegraphics[width=2.0\columnwidth,angle=0]{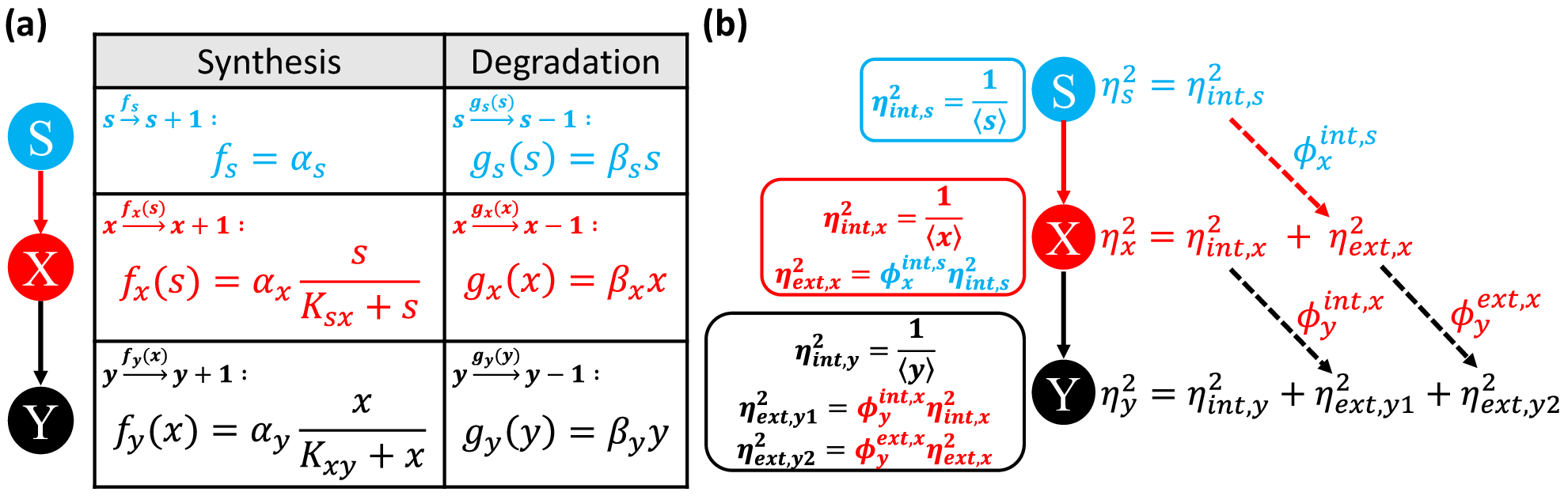}
\caption{(color online) (a) Schematic of the kinetics of a generic two-step cascade. The synthesis and degradation of S is denoted by $f_s=\alpha_s$ and $g_s(s)=\beta_s s$, respectively, and resembles a simple birth-death process. $\alpha_s$ and $\beta_s$ are synthesis and degradation rate constants, with unit (molecules/V)min$^{-1}$ and min$^{-1}$, respectively. Synthesis of X and Y is characterized by $s/(K_{sx}+s)$ and $x/(K_{xy}+x)$, respectively, that captures the binding of the upstream regulator to the promoter of the downstream gene \cite{Alon2006}. $K_{ij}$ refers to the binding affinity of the upstream protein $i$ to the promoter of the downstream gene $j$. The binding affinity $K_{ij}$ (expressed in molecules/V) defines a threshold in the concentration of $i$ necessary to express the gene product $j$ significantly \cite{Alon2006}. The rate constants $\alpha_x$ and $\alpha_y$ with unit (molecules/V) min$^{-1}$ take care of the synthesis of X and Y, respectively. The degradation of X and Y are linear functions with rate constants $\beta_x$ and $\beta_y$, respectively, with unit min$^{-1}$. In the text, $s$, $x$, and $y$ account for the copy number/volume of S, X, and Y, respectively.
(b) Schematic of noise propagation from S to Y via X. The analytical expression of various intrinsic and extrinsic noise components is shown in Appendix~C. Here, $\langle s\rangle$, $\langle x\rangle$, and $\langle y\rangle$ stand for the mean copy number of S, X, and Y, respectively, at steady state.
}
\label{f1}
\end{figure*}

To address noise propagation in a cascade of gene expression, we consider the generic two-step cascade (TSC) S$\rightarrow$X$\rightarrow$Y representing gene regulatory network (GRN). Here, S, a transcription factor (TF), regulates protein X's production, which regulates protein Y's output. For associated kinetics and corresponding master equation, we refer to  Fig.~\ref{f1}a and Eq.~(\ref{S1}), respectively. TSC is a recurring architecture frequently found in more complex networks, often associated with additional control mechanisms \cite{Alon2006,Ferrell2022}. 
Recent experimental investigation in synthetic transcriptional cascades having multiple stages of repression highlights the ability to exhibit ultrasensitive responses and low-pass filtering, resulting in robust behavior in input fluctuations \cite{Hooshangi2005}. Furthermore, noise properties in steady-state using a framework of two-stage transcriptional cascades were examined \cite{Blake2003,Bruggeman2009,Maity2015}.

Earlier studies on TSC addressed signal sensitivity and transmission using the noise decomposition framework \cite{Bruggeman2009}. Positive autoregulation at the middle node X enhances the gain-to-noise ratio compared to a simple cascade with no autoregulation \cite{Ronde2010}. The role of time scales of the constituent nodes has been shown to regulate noise and information propagation in a TSC \cite{Maity2015}. The connection between fluctuations of system components and information transmission in a TSC has been reported using the framework of partial information decomposition \cite{Williams2010}. The redundant information has been shown to increase the signal-to-noise ratio in the system \cite{Biswas2016}.

The present communication characterizes the noise propagation mechanism using the principle of noise decomposition, viz., intrinsic and extrinsic noise. The intrinsic noise is an inherent property of the system, whereas the extrinsic noise is due to environmental fluctuations. The method of noise decomposition was first experimentally shown using a dual reporter assay \cite{Elowitz2002} with a subsequent rigorous theoretical formalism \cite{Swain2002}. It is essential to mention that the principle of noise decomposition is equivalent to the Law of Total Variance \cite{Ross2014}. The noise decomposition principle has been further extended to analyze dynamic noise \cite{Hilfinger2011} and generalization to higher moments \cite{Hilfinger2012}. However, the characteristic distinction between intrinsic and extrinsic noise depends on the definition of the system \cite{Paulsson2004,Paulsson2005}. In a TSC, the intrinsic noise of each component (S, X, and Y) is due to low copies of gene products, i.e., proteins. On the other hand, the extrinsic noise in an element, say, X, is accumulated due to noise coming from S. In this context, S is an extrinsic variable (``environment") for X. Similarly, for Y, S and X both act as extrinsic variables (Fig.~\ref{f1}b).


Investigating the role of intrinsic and extrinsic noise components in cellular functions, both in terms of their drawbacks  \cite{Kaern2005} and advantages \cite{Rao2002,Eldar2010,Herranz2010,Tsimring2014}, is therefore of great significance. Previous studies have suggested that both noise components (intrinsic and extrinsic) can act as degrading factors in information transmission in cellular responses \cite{Tkacik2008a,Cheong2011,Uda2013,Selimkhanov2014,Voliotis2014,Garner2017,Potter2017,Suderman2017,Granados2018}. However, a recent communication shows enhancement of information transmission due to cell-to-cell variability (extrinsic noise in this context) in skeletal muscle \cite{Wada2020}. Mutual information (MI), a measure of statistical dependency between two variables in information theory, quantifies information-transmission capacity \cite{Shannon1948,Shannon1963,Cover1991}.

Regulatory cascades in biological systems have long been recognized for their critical role in transmitting information and orchestrating various cellular processes. However, the extent to which intrinsic and extrinsic noise contribute to the transmission of information within these cascades remains relatively unexplored. The present study addresses this gap by considering the intrinsic and extrinsic noise components and decomposing the extrinsic noise of the final gene product Y  (Fig.~\ref{f1}b). Our analysis identifies distinct noise processing channels that govern noise propagation along the cascade. We also delve into the impact of each decomposed noise term on information transmission along the cascade. Despite the apparent simplicity of the architecture, the TSC reveals an intricate relationship between noise and information, offering valuable insights into the underlying mechanisms that govern cellular functions and dynamics.

The manuscript is structured as follows: Section~\ref{s2} discusses the mathematical tools used to analyze the noise decomposition of the total output noise, revealing distinct noise processing channels between the nodes of TSC. In Section~\ref{s3}, we explore the impact of intrinsic and extrinsic noise components on information transmission across the TSC. We observe a sequential hierarchy within the intrinsic noise of the nodes of TSC, where the cascade achieves optimal information transmission from input to output. Finally, Section~\ref{s4} summarizes the key findings and concludes.


\section{The model and methods}
\label{s2}

The stochastic kinetics of a generic TSC is given by
\begin{subequations}
\begin{eqnarray}
s \stackrel{f_s}{\longrightarrow} s+1, \; \; \;
x \stackrel{f_x (s)}{\longrightarrow} x+1, \; \; \;
y \stackrel{f_y (x)}{\longrightarrow} y+1, \\
s \stackrel{g_s (s)}{\longrightarrow} s-1, \; \; \;
x \stackrel{g_x (x)}{\longrightarrow}  x-1, \; \; \;
y \stackrel{g_y (y)}{\longrightarrow}  y-1,
\end{eqnarray}
\end{subequations}

\noindent where $s$, $x$, and $y$ stand for the copy number/volume of S, X, and Y, respectively. Here $f_i$ and $g_i$ ($i \equiv s,x,y$) correspond to the synthesis and degradation functions of the system components, respectively. The explicit expressions of $f_i$ and $g_i$ are outlined in Fig.~1a. 
The master equation following the stochastic kinetics outlined in Eq.~(1) is \cite{Gardiner2009}
\begin{eqnarray}
\label{S1}
\frac{dP(s,x,y;t)}{dt} & = & \sum_{i=s,x,y} \left[
(\mathbb{E}^{+1}_i - 1) g_i P(s,x,y;t) \right. \nonumber \\
&& \left. + (\mathbb{E}^{-1}_i - 1) f_i P(s,x,y;t) \right].
\end{eqnarray}

\noindent  In the above equation, $\mathbb{E}^{\pm}_i$ refers to the step operator which either step-up ($\mathbb{E}^{+1}_i$) or step-down ($\mathbb{E}^{-1}_i$) the copy numbers of the respective components by unity. We employ van Kampen's system size expansion \cite{Kampen2007} to derive the Lyapunov equation (see Appendix~A), which provides the statistical moments associated with the TSC (see Appendix~B).

 \subsection{Noise decomposition in a two-step cascade}

In a GRN, the noise associated with the $i$-th node is measured by the square of the coefficient of variation, $\eta_i^2 = \sigma_i^2/\langle i \rangle^2$. 
where $\sigma_i^2$ and $\langle i \rangle$ refer to the variance and the mean copy number, respectively, at steady state.
As per the noise decomposition formalism $\eta_i^2$ can be written as $\eta_i^2=\eta_{int,i}^2+\eta_{ext,i}^2$ \cite{Swain2002,Paulsson2005,Hilfinger2011} where $\eta_{int,i}^2$ and $\eta_{ext,i}^2$ are intrinsic and extrinsic noise, respectively. The intrinsic noise $\eta_{int,i}^2$ arises due to the birth-death processes, whereas the extrinsic noise $\eta_{ext, i}^2$ is fed from its upstream regulator.
The noise associated with the nodes S, X, and Y of a TSC is (see Appendix~C)
\begin{eqnarray}
\label{eqn2}
\eta_s^2 &=& \eta_{int,s}^2, \\
\eta_x^2 &=& \eta_{int,x}^2 + \eta_{ext,x}^2,
\label{eqn3} \\
\eta_y^2 &=& \eta_{int,y}^2 + \eta_{ext,y1}^2 +\eta_{ext,y2}^2.
\label{eqn4}
\end{eqnarray}

\noindent In Eq.~(\ref{eqn2}), S contains only intrinsic noise ($\eta_{int,s}^2$) due to the Poisson kinetics. The noise associated with X, however, contains both intrinsic ($\eta_{int,x}^2$) and extrinsic parts ($\eta_{ext,x}^2$). The extrinsic noise $\eta_{ext,x}^2$ arises due to the propagation of noise from S to X, \textit{i.e.}, from $\eta_{int,s}^2$. Similarly, the total noise of Y, $\eta_y^2$, has intrinsic noise $\eta_{int,y}^2$ and a summation of two extrinsic noises $\eta_{ext,y1}^2$ and $\eta_{ext,y2}^2$. The source of extrinsic noise $\eta_{ext,y1}^2$ is the intrinsic noise $\eta_{int,x}^2$. On the other hand, $\eta_{ext,y2}^2$ is generated due to the contribution of $\eta_{ext,x}^2$. We refer to Fig.~\ref{f1}b for the noise flow along the cascade. The general noise decomposition technique can be extended to a linear cascade with nodes $\geqslant 3$.

The explicit expressions of the extrinsic noise given in Eqs.~(\ref{eqn2}-\ref{eqn4}) are (see Appendix~C)
\begin{eqnarray}
\label{eqn5}
\eta_{ext,x}^2 &=& \phi_x^{int,s} \eta_{int,s}^2, \\
\label{eqn6}
\eta_{ext,y1}^2 &=& \phi_y^{int,x} \eta_{int,x}^2, \\
\label{eqn7}
\eta_{ext,y2}^2 &=& \phi_y^{ext,x} \eta_{ext,x}^2 = \phi_y^{ext,x} \phi_x^{int,s} \eta_{int,s}^2,
\end{eqnarray}

\noindent where,
\begin{eqnarray}
\label{eqn8}
\phi_x^{int,s} &=& \tau_{x,s}
\left( \frac{K_{sx}}{K_{sx}+\langle s\rangle} \right)^2, \\
\label{eqn9}
\phi_y^{int,x} &=& \tau_{y,x}
\left( \frac{K_{xy}}{K_{xy}+\langle x\rangle} \right)^2, \\
\label{eqn10}
\phi_y^{ext,x} &=& \tau_{sy,x}^{-1} \phi_y^{int,x}.
\end{eqnarray}

\noindent Eqs.~(\ref{eqn5}-\ref{eqn7}) suggest that the quantity $\phi$ measures the fraction of upstream noise propagated to its corresponding downstream node. Depending on the cascade architecture and the model parameters, $\phi$ quantitates the pool of noise that flows downstream (see Eqs.~(\ref{eqn8}-\ref{eqn10})). To be specific, $\phi_x^{int,s}$ quantifies the fraction of intrinsic noise of S transmitted to X that builds up the extrinsic noise pool of X, $\eta_{ext,x}^2$ (Fig.~\ref{f1}b). $\phi_y^{int,x}$ measures the fraction of intrinsic noise of X propagated to Y to generate the extrinsic noise pool $\eta_{ext,y1}^2$ of Y (Fig.~\ref{f1}b). Similarly, $\phi_y^{ext,x}$ measures the fraction of extrinsic noise of X that flows down to Y and generates the second part of the extrinsic noise pool $\eta_{ext,y2}^2$ of Y (Fig.~\ref{f1}b). The measure $\phi$, thus, characterizes the noise transmission channels between different nodes of the cascade and identifies how noise propagation influences various species of the linear cascade. It is essential to mention that $\phi$ always remains $< 1$ independent of different model parameters.


\begin{figure*}[t]
\includegraphics[width=2.0\columnwidth,angle=0]{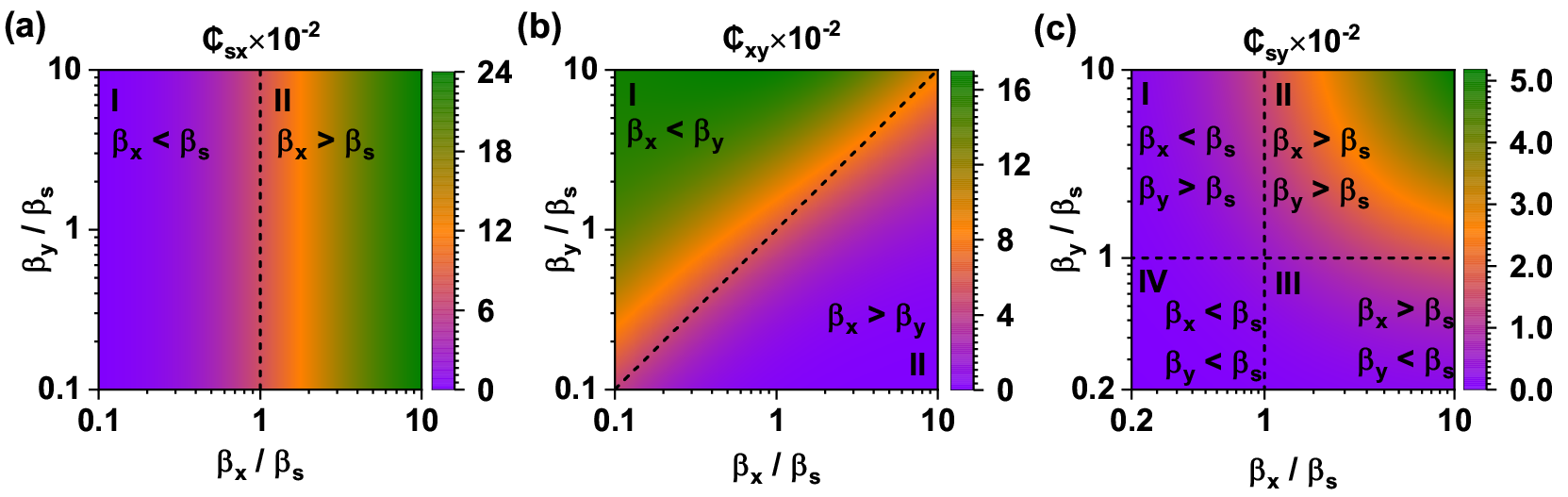}
\caption{(color online) Contour map of (a) $\mathbb{C}_{sx}$, (b) $\mathbb{C}_{xy}$, and (c) $\mathbb{C}_{sy}$ as functions of $\beta_x/\beta_s$ and $\beta_y/\beta_s$. The parameters used to generate the maps are $\langle s\rangle=50$ molecules/V, $\langle x\rangle=100$ molecules/V, $\langle y\rangle=100$ molecules/V, $K_{sx}=\langle s\rangle$, and $K_{xy}=\langle x\rangle$.
}
\label{f2}
\end{figure*}

In Eqs.~(\ref{eqn8}-\ref{eqn10}), $\tau$-s are the scaled time scale \cite{Paulsson2004,Paulsson2005} defined as $\tau_{i,j}:=\beta_i/(\beta_i+\beta_j)$ and $\tau_{ij,k}:=(\beta_i+\beta_j)/(\beta_i+\beta_j+\beta_k)$ where $\beta_i$ ($i \in \{s,x,y \}$) corresponds to the degradation rate constant of the $i$-th species. The expressions of $\tau$-s indicate that $\tau_{x,s}<1$, $\tau_{y,x}<1$, and $\tau_{sy,x} < 1$. Moreover, the second factors in Eqs.~(\ref{eqn8},\ref{eqn9}) are also less than unity. This results in $\phi_x^{int,s}<1$ and $\phi_y^{int,x}<1$. On the other hand, although $\tau_{sy,x}^{-1}>1$, we assume that this factor will not overpower $\phi_y^{int,x}$, which is much less than one, and hence leads to $\phi_y^{ext,x} < 1$. This assumption remains valid as long as the separation of degradation time scale is maintained, i.e., $\beta_s < \beta_x < \beta_y$. The usage of the separation of time scale adopted in the present work is motivated by earlier communications \cite{Bruggeman2009, Maity2015} where the authors detailedly studied noise and information propagation in several biochemical motifs. The theoretical analysis of Bruggeman et al. \cite{Bruggeman2009} and Maity et al. \cite{Maity2015} show that the separation of degradation time scale maintains maximum noise propagation along the cascade.

The noise propagation from S to X opens up a single noise propagation channel characterized by $\phi_x^{int,s}$. On the other hand, noise transmission from X to Y is regulated by two different channels $\phi_y^{int,x}$ and $\phi_y^{ext,x}$. Noise flow from S to Y via X is characterized by $\phi_x^{int,s}$ and $\phi_y^{ext,x}$. We note that an increase in cascade length opens up multiple new channels of noise propagation. For example, in a cascade S$\rightarrow$X$\rightarrow$Y$\rightarrow$Z, additional downstream channels open up for noise propagation from Y to Z.

To substantiate our theoretical results, we employ stochastic simulation algorithm \cite{Gillespie1976, Gillespie1977} to generate numerical data. To this end, we use the biochemical kinetics shown in Fig.~\ref{f1}a for simulation. Each numerical data is the mean of $10^6$ independent realizations. The unit of mean copy number is molecules/V, where V is the unit of cellular volume and the unit of $\beta_i$ ($i \in s,x, ~\text{and}~y$) is min$^{-1}$.


\section{Results and discussion}
\label{s3}

The notion of mutual information \cite{Shannon1948, Shannon1963} is used to address the interplay between noise and information transmission. The mutual information between two random variables $i$ and $j$ is written as $I(j;i) := \sum_{i,j} p(i,j) \log_2 [p(i,j)/p(i)p(j)]$, where $p(i)$ and $p(j)$ are marginal probability distributions and $p(i,j)$ refer to the joint probability distribution of the variables \cite{Cover1991}. For a bivariate system $i$ and $j$ obeying Gaussian statistics, the channel capacity is $\mathbb{C}_{ij} := I(i;j)=(1/2)\log_2[1+\mathbb{S}_{ij}]$, where $\mathbb{S}_{ij}$ ($=\eta_{ij}^4/(\eta_i^2 \eta_{j|i}^2)$) is the signal-to-noise ratio (SNR) of the respective channels \cite{Tostevin2010}. Here $\eta_{ji}^2$, $\eta_i^2$, and $\eta_{j|i}^2$ stand for the normalised covariance ($\sigma_{ji}^2/(\langle j\rangle \langle i\rangle)$), variance $(\sigma_i^2/\langle i\rangle^2$) and conditional variance ($\sigma_{j|i}^2/\langle j\rangle^2$), respectively (see Appendix~D for the explicit expressions of $\mathbb{S}_{ij}$ and $\mathbb{C}_{ij}$). In the present scenario, we have three possible information processing channels, i.e., S$\rightarrow$X, X$\rightarrow$Y, and the overall channel S$\rightarrow$X$\rightarrow$Y. We note that, for the Gaussian channel, channel capacity is a function of SNR and equivalent to the MI based on the statistical moments of the system components \cite{Cover1991, Mackay2003, Tostevin2010}. However, for non-Gaussian systems, MI computed from moments yields a lower bound on channel capacity \cite{Mitra2001}. A probabilistic framework is necessary to determine exact channel capacity in non-Gaussian scenarios, where channel capacity is obtained by maximizing MI over all possible input distributions \cite{Cover1991, Mackay2003}. In the following, we analyze our findings using channel capacity to align with the channel-based analysis of the TSC.

\subsection{Hierarchy of relaxation time scales determines efficient information transmission}

To understand how the relaxation time scales regulate information transmission across different components of TSC, we show the channel capacities $\mathbb{C}_{ij}$ as functions of $\beta_x/\beta_s$ and $\beta_y/\beta_s$ (Fig.~\ref{f2}).
The transmission of information from S to X (measured by $\mathbb{C}_{sx}$) is enhanced for $\beta_x > \beta_s$ (regime II in Fig.~\ref{f2}a). This inequality points out that input signal S fluctuates on a relatively slower time scale than X. As a result, the middle node X effectively captures the slower variations in S. Conversely, when $\beta_x < \beta_s$, X fluctuates more slowly than S. Consequently, the slower varying X cannot accurately sense the faster variations in S, leading to reduced information transmission (regime I in Fig.~\ref{f2}a).

On a similar note, the information transmission from X to Y (measured by $\mathbb{C}_{xy}$) is amplified in the regime where $\beta_x < \beta_y$ (regime~I in Fig.~\ref{f2}b). This scenario allows X to fluctuate on a slower time scale than Y and enables Y to perceive the slow variations in X accurately. On the other hand, for $\beta_x > \beta_y$, Y fluctuates relatively slower compared to X and fails to effectively capture the faster variations in X, resulting in a notable reduction in information transmission (regime~II in Fig.~\ref{f2}b).


\begin{figure*}[t]
\includegraphics[width=2.0\columnwidth,angle=0]{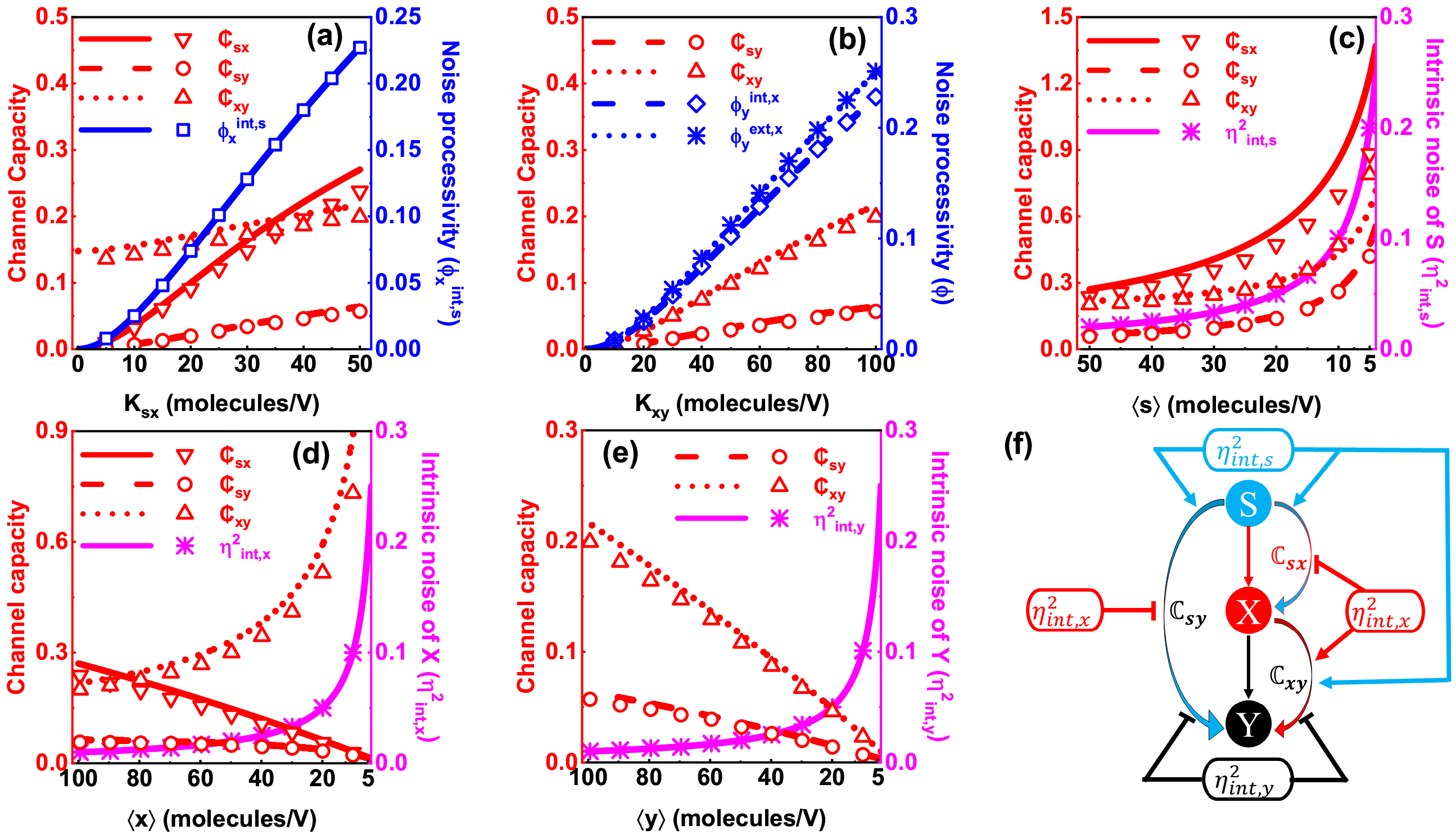}
\caption{(color online) (a,b) Variations in channel capacity and noise processivity $\phi$ as a function of $K_{sx}$ and $K_{xy}$. The parameters used are $\beta_s=0.1$ min$^{-1}$, $\beta_x=1.0$ min$^{-1}$, $\beta_y=10.0$ min$^{-1}$, $\langle s\rangle=50$ molecules/V, $\langle x\rangle=100$ molecules/V, and $\langle y\rangle=100$ molecules/V. For (a) $K_{xy}=\langle x\rangle$ and for (b) $K_{sx}=\langle s\rangle$.
(c-e) Variations in channel capacity and intrinsic noise of S, X, and Y as a function of mean copy number. The parameters used are $\beta_s=0.1$ min$^{-1}$, $\beta_x=1.0$ min$^{-1}$, $\beta_y=10.0$ min$^{-1}$, $K_{sx}=\langle s\rangle$, and $K_{xy}=\langle x\rangle$. For (c) $\langle x\rangle=100$ molecules/V and $\langle y\rangle=100$ molecules/V. For (d) $\langle s\rangle=50$ molecules/V and $\langle y\rangle=100$ molecules/V, and for (e) $\langle s\rangle=50$ molecules/V and $\langle x\rangle=100$ molecules/V.
In (a-e), the lines are due to theoretical expression, and the symbols are generated by numerical simulation using stochastic simulation algorithm \cite{Gillespie1976,Gillespie1977}.
(f) Schematic presentation of channel capacities in a TSC and the role of intrinsic noise on information propagation. The pointed ($\rightarrow$) and blunt ($\dashv$) arrowheads stand for activation and repression, respectively.
}
\label{f3}
\end{figure*}

For overall information transmission from S to Y via X (measured by $\mathbb{C}_{sy}$), we observe amplification of $\mathbb{C}_{sy}$ when $\beta_x > \beta_s$ and $\beta_y > \beta_s$ (regime II in Fig.~\ref{f2}c). These inequalities indicate that the signal S fluctuates at a slower rate compared to both X and Y. As a result, Y can effectively capture the slow fluctuations present in S via X, leading to a high degree of information transmission from S to Y. However, in other domains of Fig.~\ref{f2}c (I, III, and IV), $\mathbb{C}_{sy}$ is minimal due to the discrepancy in maintaining a sequential hierarchy of fluctuations among the components S, X, and Y. Such discrepancies lead to inefficient information transmission and contrast with the optimal conditions observed in regime II.

The previous discussion on various regimes with a high level of information flow (regimes II in Fig.~\ref{f2}a, I in Fig.~\ref{f2}b, and II in Fig.~\ref{f2}c) reveals a clear hierarchy of the relaxation time scales, i.e., $\beta_s<\beta_x<\beta_y$ which ensures that the cascade operates at its peak efficiency in terms of information transmission. By maintaining this temporal hierarchy, the cascade can effectively capture and transmit information across its nodes, optimizing its functionality. It is important to note that Bruggeman et al. \cite{Bruggeman2009} and Maity et al. \cite{Maity2015} reported a similar kind of analysis to show the role of relaxation time scales on noise transmission and information transmission in a TSC.

\subsection{Impact of intrinsic and extrinsic noise on information transmission}

For $\beta_s < \beta_x < \beta_y$ the scaled time scales $\tau$-s become approximately equal to 1  (see Appendix~D). In this limit, the analytical expressions of channel capacities are,
\begin{eqnarray}
\mathbb{C}_{sx} &=& \frac{1}{2} \log_2 \left[
1 + \frac{\phi_x^{int,s} \eta_{int,s}^2}{\eta_{int,x}^2}
\right], \label{eqn18} \\
\mathbb{C}_{sy} &=& \frac{1}{2} \log_2 \left[
1 + 
\frac{\phi_x^{int,s} \phi_y^{ext,x} \eta_{int,s}^2}
{\eta_{int,y}^2 + \phi_y^{int,x} \eta_{int,x}^2}
\right], \label{eqn19} \\
\mathbb{C}_{xy} &=& \frac{1}{2}
\log_2 \left[ 1+
\frac{A}{B} \right], 
\label{eqn20}
\end{eqnarray}

\noindent where
\begin{eqnarray*}
A &=& \phi_y^{int,x}\eta_{int,x}^4 + (\phi_x^{int,s} \eta_{int,s}^2 
+ 2 \eta_{int,x}^2) \phi_x^{int,s} \phi_y^{ext,x} \eta_{int,s}^2, \\
B &=& \eta_{int,x}^2 \eta_{int,y}^2 + (\eta_{int,y}^2+\phi_y^{int,x}\eta_{int,x}^2) \phi_x^{int,s} \eta_{int,s}^2 \nonumber \\
&& - \phi_x^{int,s} \phi_y^{ext,x} \eta_{int,s}^2 \eta_{int,x}^2.
\end{eqnarray*}

As the input signal S acts as an extrinsic variable for both X and Y, increase in $\phi_x^{int,s}$ amplifies all three channel capacities, $\mathbb{C}_{sx}$, $\mathbb{C}_{sy}$, and $\mathbb{C}_{xy}$ (see Eqs.~(\ref{eqn18}-\ref{eqn20})). Notably, $\mathbb{C}_{sx}$ exhibits the most rapid increase in response to variations in $\phi_x^{int,s}$ compared to $\mathbb{C}_{sy}$ and $\mathbb{C}_{xy}$ (Fig.~\ref{f3}a). On the other hand, for X$\rightarrow$Y channel, X acts as an extrinsic variable for Y. Therefore, an increase in $\phi_y^{int,x}$ and $\phi_y^{ext,x}$ leads to enhancement of the channel capacities $\mathbb{C}_{sy}$ and $\mathbb{C}_{xy}$ (see Eqs.~(\ref{eqn19},\ref{eqn20}) and Fig.~\ref{f3}b). These findings illustrate that the magnitude of $\phi$-s, which characterizes noise propagation along each channel, is closely associated with the information transmission capacity of each channel. $\phi$ thus indicates the noise processivity of individual noise processing channels along the regulatory edges. By `processivity', we mean the ability to 
transduce the noise along a particular channel.

The noise processivity $\phi$ generates an extrinsic noise pool at the downstream node of each channel, which enhances information transmission along the respective channel. Therefore, the contribution of extrinsic noise at the downstream node positively facilitates information transmission from an upstream node.
Eq.~(\ref{eqn18}-\ref{eqn20}) further reveals that the intrinsic noise $\eta_{int,s}^2$ enhances all the three channel capacities (Fig.~\ref{f3}c,f). The input noise $\eta_{int,s}^2$ thus facilitates information transmission downstream as it acts as an extrinsic variable for X and Y. In other words, the more the magnitude of $\eta_{int,s}^2$, the more the downstream nodes X and Y sense the input noise as information.

The influence of the intrinsic noise of X ($\eta_{int,x}^2$) on $\mathbb{C}_{sx}$ and $\mathbb{C}_{sy}$ is repressing in nature (Fig.~\ref{f3}d,f). The increase in intrinsic noise of X acts as a constraint, limiting X's ability to accurately detect the signal S. Consequently, the information transmission from S to X ($\mathbb{C}_{sx}$) diminishes. The information transmission from S to Y ($\mathbb{C}_{sy}$) also diminishes as Y gains information from S via X. The information transmission from X to Y ($\mathbb{C}_{xy}$) increases due to $\eta_{int,x}^2$ (Fig.~\ref{f3}d,f) as $\eta_{int,x}^2$ acts as an extrinsic variable for the channel X$\rightarrow$Y. The intrinsic noise of Y,  $\eta_{int,y}^2$, increases the variability of Y. As a consequence, $\eta_{int,y}^2$ acts as a potent limiting factor for which the information gained by Y from both S and X ($\mathbb{C}_{sy}$ and $\mathbb{C}_{xy}$) is reduced (Fig.~\ref{f3}e,f).

\subsection{Sequential hierarchy in intrinsic noises regulates information transmission}


\begin{figure}[t]
\includegraphics[width=1.0\columnwidth,angle=0]{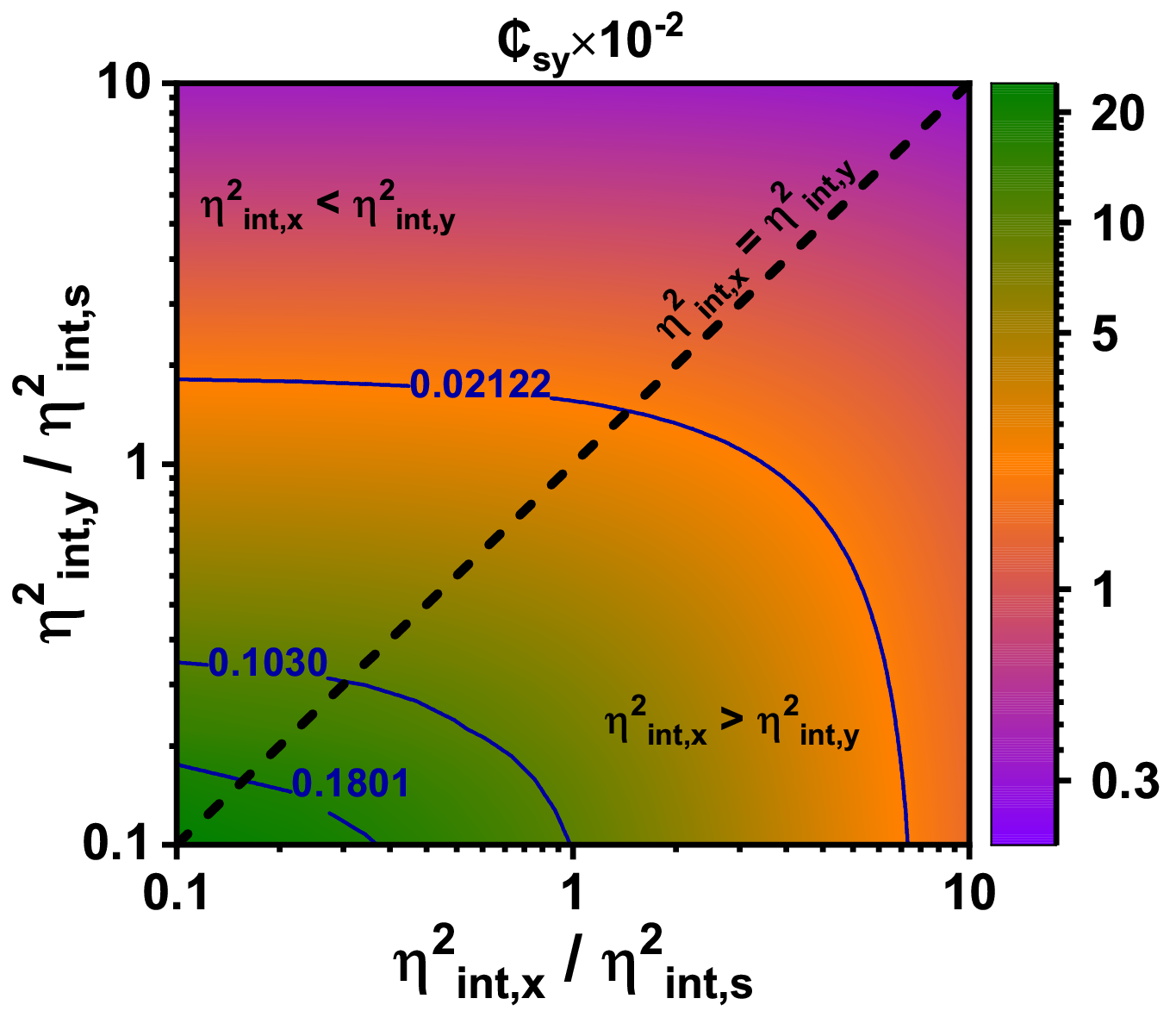}
\caption{(color online) Contour map of $\mathbb{C}_{sy}$ as functions of $\eta_{int,x}^2/\eta_{int,s}^2$ and $\eta_{int,y}^2/\eta_{int,s}^2$. The parameters used to generate the maps are $\beta_s=0.1$ min$^{-1}$, $\beta_x=1.0$ min$^{-1}$, $\beta_y=10.0$ min$^{-1}$, $\langle s\rangle=50$ molecules/V, $K_{sx}=\langle s\rangle$, and $K_{xy}=\langle x\rangle$. The variations in $\eta_{int,x}^2$, and $\eta_{int,y}^2$ are obtained by tuning $\langle x\rangle$ and $\langle y\rangle$, respectively using the relations $\eta_{int,x}^2 = 1/\langle x\rangle$, and $\eta_{int,y}^2 = 1/\langle y\rangle$.
}
\label{f4}
\end{figure}

The preceding discussion suggests that while the noise processing capacities ($\phi$-s) always facilitate information transmissions across different regulatory channels, the intrinsic noise of S, X, and Y play distinct roles in this process  (Fig.~\ref{f3}f). We now focus on the overall information transmission $\mathbb{C}_{sy}$, to examine how the three intrinsic noise components $\eta_{int,s}^2$, $\eta_{int,x}^2$, and $\eta_{int,y}^2$ jointly regulate information transmission from input S to output Y via X. We opted not to analyze the remaining information transmission measures (i.e., $\mathbb{C}_{sx}$ and $\mathbb{C}_{xy}$) as these terms individually do not offer a complete insight into how the three intrinsic noises collectively regulate information transmission across the entire system.

To this end we show $\mathbb{C}_{sy}$  as functions of $\eta_{int,x}^2/\eta_{int,s}^2$ and $\eta_{int,y}^2/\eta_{int,s}^2$ in Fig.~\ref{f4}. We identify a specific regime characterized by $(\eta_{int,x}^2/\eta_{int,s}^2)<1$ and $(\eta_{int,y}^2/\eta_{int,s}^2)<1$, where maximum information transmission from S to Y occurs. In this regime, the higher intrinsic noise of S and lower intrinsic noise of X and Y facilitate efficient information transfer from S to Y (Fig.~\ref{f3}c-\ref{f3}f). Along the diagonal line in Fig.~\ref{f4} $\eta_{int,x}^2 = \eta_{int,y}^2$ below which $\eta_{int,x}^2 > \eta_{int,y}^2$, while above it, $\eta_{int,x}^2 < \eta_{int,y}^2$. The maximum information transmission domain predominantly lies in the lower regime, i.e., where $\eta_{int,x}^2 > \eta_{int,y}^2$. Therefore, maintenance of $\eta_{int,x}^2 > \eta_{int,y}^2$ is the preferred configuration to achieve maximum information transfer across the cascade. This analysis reveals that maximum information transmission across a TSC can be achieved when the three intrinsic noise components follow the relationship $\eta_{int,s}^2 > \eta_{int,x}^2 > \eta_{int,y}^2$.

Our analysis unveils a pivotal hierarchy among various intrinsic noises, where the intrinsic noise levels decrease sequentially from S to Y. This hierarchy is essential for achieving maximum information transfer from S to Y in a TSC. By maintaining this hierarchy, signal transduction networks can optimally process and transmit information. This reveals the intricate mechanisms by which cellular systems regulate information flow, underscoring the significance of intrinsic noise modulation in cellular communication and signaling pathways.


\section{Conclusion}
\label{s4}

We presented a theoretical analysis of information processing in a TSC within the purview of noise decomposition. We demonstrated the presence of distinct noise processing channels between the system components S, X, and Y. The extent of noise propagation along these channels depends on the noise processivity $\phi$. The intrinsic noise of the signal $\eta_{int,s}^2$, together with noise processivity ($\phi$), acts as extrinsic variables and alleviates the overall information transmission along the cascade. On the other hand, the intrinsic noise of X, $\eta_{int,x}^2$ put a limit in information transmission along the channels S$\rightarrow$X ($\mathbb{C}_{sx}$) and S$\rightarrow$X$\rightarrow$Y ($\mathbb{C}_{sy}$) and thereby acts as a bottleneck that reduces information transfer capacities. On the contrary, $\eta_{int,x}^2$ enhances information transmission along X$\rightarrow$Y ($\mathbb{C}_{xy}$). Similarly, intrinsic noise of Y, $\eta_{int,y}^2$ hinders information flow along the channels S$\rightarrow$X$\rightarrow$Y ($\mathbb{C}_{sx}$) and X$\rightarrow$Y ($\mathbb{C}_{xy}$).  Moreover, while analyzing the overall information transmission across the TSC, a sequential hierarchy emerges among the intrinsic noise levels of S, X, and Y. This hierarchy facilitates optimal information transmission from input S to output Y. In the hierarchy, S exhibits the highest intrinsic noise, followed by X and then Y, i.e., $\eta_{int,s}^2 > \eta_{int,x}^2 > \eta_{int,y}^2$. As long as this hierarchy is maintained within the components of the TSC, the cascade allows the highest information transmission from input S to output Y.

While our analysis focused on analytical and computational predictions, direct experimental evidence for the hierarchy in intrinsic noise levels is currently lacking. However, verification can be achieved by constructing a synthetic gene circuit of a TSC using recombinant technology. Fluorescent reporter assays of the gene circuit will be useful to quantify the steady-state expression levels associated with each component (S, X, and Y). Subsequently, the variance of each component's expression level can be calculated from the reporter assay data. A data-driven decomposition of the variance will be useful using the modified total variance decomposition framework proposed by Hiffinger et al. \cite{Hilfinger2011}. The decomposition formalism applied to the fluorescent assay of output Y results in,
\begin{eqnarray*}
\eta_y^2 &=& \frac{\sigma_y^2}{\langle y\rangle^2}
= \underbrace{
\frac{\langle \sigma_{y|\mathbf{s,x}}^2 \rangle}{\langle y\rangle^2}
}_{\rm intrinsic}
+ \underbrace{
\frac{\sigma_{\langle y|\mathbf{s,x}\rangle}^2}{\langle y\rangle^2}
}_{\rm extrinsic},
\end{eqnarray*}

\noindent where $\mathbf{s}=\{ s_t, s_{t-1} \}$ and $\mathbf{x}=\{ x_t, x_{t-1} \}$, and $t$ accounts for the steady state time. In the above equation, $\sigma_y^2$, $\langle \sigma_{y|\mathbf{s,x}}^2\rangle$, and $\sigma_{\langle y|\mathbf{s,x}\rangle}^2$ stand for the variance of Y,  ensemble average of conditional variance of Y and variance of conditional mean of Y, respectively.  Moreover, $\langle y\rangle$ denotes the mean of Y reporter over many realizations.
Hiffinger's framework suggests the utilization of the entire history of S and X for decomposition. However, we have modeled the variables S, X, and Y as Markov chains. In a Markov chain, the future state depends only on the present state, and knowledge of the entire history provides no additional information. Therefore, for the decomposition, we only need the variables' current state ($t$) and the immediate past state ($t-1$), simplifying Hiffinger's framework without compromising the decomposition's validity. This allows for a more practical implementation in the experimental setting. The decomposition allows for the separation of intrinsic and extrinsic noise contributions for each variable. We note that the decomposition was initially formulated based on the dual reporter method \cite{Elowitz2002, Swain2002}. In another study, noise decomposition from experimental data was also performed to quantify the pathway-delineated noise components in gene expression \cite{Ham2021}.

In summary, the role of noise processivity, $\phi$, is straightforward. It consistently enhances information transmission across the TSC as $\phi$ essentially captures the extent of extrinsic noise propagation. Additionally, the intrinsic noise of S contributes to alleviating information transmissions along each channel, whereas the intrinsic noise of Y hinders the same. However, the role of the intrinsic noise of X is intricate. It alleviates when information is transmitted from X but suppresses when information is sent to X and through X to Y. In this context, the hierarchy of intrinsic noises plays a significant role in preserving the highest information transmission from S to Y. Thus, noise processivity and intrinsic noises modulate the information transmission capacity of the TSC.

The interplay between intrinsic and extrinsic noise constitutes an optimum signal transduction machinery. When there is a need to precisely sense and detect the signal for adaptability, the genes within the cascade exhibit dynamics in which the molecular intrinsic fluctuations become negligible. This ensures that the system's inherent noise does not hinder the accurate reception of the signal. Conversely, when the signal is detrimental to the cellular response, the cascade's gene expression dynamics adapt to increase the intrinsic noise of the gene products. Such a strategy can be seen as a protective mechanism, where the system becomes less responsive to external fluctuations, thereby prioritizing robustness that might otherwise disrupt its normal functioning.

 
\begin{acknowledgments}
MN thanks SERB, India, for the National Post-Doctoral Fellowship [PDF/2022/001807]. 
\end{acknowledgments}


\appendix

\section{System size expansion and Lyapunov equation}

We use the notation $\mathbf{N}$ to represent a three-dimensional vector of copy numbers of the components, i.e., $\mathbf{N} \in \{ n_s \equiv s, n_x \equiv x, n_y \equiv y \}$.
We rewrite Eq.~(\ref{S1}) in terms of macroscopic concentration of each component defined as $c_i := \lim\limits_{\Omega \to \infty} (n_i/\Omega)$ with $i \in \{s, x, y\}$ \cite{Hayot2004,Kampen2007} where $\Omega$ is the reaction volume,
\begin{eqnarray}
\label{S2}
\frac{dP(\pmb{N};t)}{dt} &=& \Omega
\sum_{i=s,x,y} \left[ (\mathbb{E}_i^{+1}-1) g_i(n_i/\Omega) P(\pmb{N};t) \right. \nonumber \\
&& \left. + (\mathbb{E}_i^{-1}-1) f_i(\pmb{N}/\Omega) P(\pmb{N};t) 
\right].
\end{eqnarray}

\noindent In the above equation, $f_i(\pmb{N}/\Omega)$ represents the generalised form of any reaction network. For TSC, it specifically translates to $f_s$, $f_x(n_s/\Omega)$, and $f_y(n_x/\Omega)$, representing the production of S, X, and Y, respectively. In the following, we provide the pedagogical formulation of van Kampen's system size expansion \cite{Kampen2007} to solve the Eq.~(\ref{S2}).
We now expand $n_i$ in $\Omega$ space of $\mathcal{O}(\Omega^{1/2})$
\begin{equation}
\label{S3}
n_i = \Omega c_i + \Omega^{1/2} \epsilon_i,
\end{equation}

\noindent where $\epsilon_i$ represents the fluctuations around $n_i$. Using Eq.~(\ref{S3}) in the left hand side of Eq.~(\ref{S2}) we have
\begin{equation}
\label{S4}
\frac{dP(\pmb{N};t)}{dt} =
\frac{d\widetilde{P}(\pmb{\epsilon};t)}{dt} - \Omega^{-1/2}
\sum_{i=s,x,y} \frac{dc_i}{dt}
\frac{d\widetilde{P}(\pmb{\epsilon};t)}{d\epsilon_i},
\end{equation}

\noindent with $\pmb{\epsilon} \in \{ \epsilon_s, \epsilon_x, \epsilon_y \}$. Eq.~(\ref{S4}) accounts for the time derivative of $P(\pmb{N};t)$ at fixed $n_i$ so that Eq.~(\ref{S3}) becomes $(d\epsilon_i/dt) = \Omega^{-1/2} (dc_i/dt)$ \cite{Elf2003,Hayot2004}.
Expansion of the step operator $\mathbb{E}^{\pm}_i$ in Eq.~(\ref{S2})  provides \cite{Elf2003,Hayot2004}
\begin{eqnarray}
\mathbb{E}_i^{+1} &=& 1 + \Omega^{-1/2} (\partial/\partial \epsilon_i)
+ (1/2) \Omega^{-1} (\partial^2/\partial \epsilon_i^2), 
\label{S5}
\\
\mathbb{E}_i^{-1} &=& 1 - \Omega^{-1/2} (\partial/\partial \epsilon_i) + (1/2) \Omega^{-1} (\partial^2/\partial \epsilon_i^2)
\label{S6}.
\end{eqnarray}

\noindent Moreover, the Taylor expansion of the synthesis and degradation functions in Eq.~(\ref{S2}) around their macroscopic values give \cite{Elf2003},
\begin{eqnarray}
\label{S7}
f_i\left(\frac{\pmb{N}}{\Omega}\right) &=& 
f_i(\pmb{c})+\Omega^{-1/2} \sum_{j=s,x,y} \epsilon_j \frac{\partial f_i(\pmb{c})}{\partial c_j},
\\
\label{S8}
g_i \left( \frac{n_i}{\Omega} \right) &=&
g_i(c_i) + \Omega^{-1/2} \epsilon_i
\frac{\partial g_i(c_i)}{\partial c_i},
\end{eqnarray}

\noindent with $\pmb{c} \in \{ c_s, c_x, c_y \}$. Inserting $P(\pmb{N};t)=\widetilde{P}(\pmb{\epsilon};t)$ and Eqs.~(\ref{S5}-\ref{S8}) into the right-hand side of Eq.~(\ref{S2}) we have
\begin{widetext}

\begin{eqnarray}
&& \frac{d\widetilde{P}(\pmb{\epsilon};t)}{dt} - \Omega^{-1/2}
\sum_{i=s,x,y} \frac{dc_i}{dt}
\frac{d\widetilde{P}(\pmb{\epsilon};t)}{d\epsilon_i}  \nonumber \\
&& = \Omega \sum_{i=s,x,y}
\left[
\left(-\Omega^{-1/2} \frac{\partial}{\partial \epsilon_i} + \frac{1}{2} \Omega^{-1} \frac{\partial^2}{\partial \epsilon_i^2} \right) 
\left(f_i(\pmb{c})+\Omega^{-1/2} \sum_{j=s,x,y} \epsilon_j \frac{\partial f_i(\pmb{c})}{\partial c_j} \right) 
\widetilde{P}(\pmb{\epsilon};t) \right. \nonumber \\
&& \left. +
\left(\Omega^{-1/2} \frac{\partial}{\partial \epsilon_i} + \frac{1}{2} \Omega^{-1} \frac{\partial^2}{\partial \epsilon_i^2} \right)
\left(g_i(c_i) + \Omega^{-1/2} \epsilon_i \frac{\partial g_i(c_i)}{\partial c_i} \right)
\widetilde{P}(\pmb{\epsilon};t)
\right].
\label{S9}
\end{eqnarray}

\end{widetext}

In the above equation, there are two dominant terms with $\mathcal{O}(\Omega^{1/2})$ and $\mathcal{O}(\Omega^{0})$. Collecting the terms of $\mathcal{O}(\Omega^{1/2})$ from both sides results in the usual deterministic rate equation for the TSC motif in terms of concentrations $c_i$,
\begin{equation}
\label{S10}
\frac{d c_i}{dt} = f_i(\pmb{c})-g_i(c_i).
\end{equation}

\noindent Using the definition of $c_i$, Eq.~(\ref{S10}) can be recast in terms of copy number $n_i$ as $dn_i/dt = f_i(\pmb{N})-g_i(n_i)$. For the components S, X, and Y, the deterministic equations become,
\begin{eqnarray}
\label{S11}
\frac{ds}{dt} &=& f_s - g_s(s), \\
\label{S12}
\frac{dx}{dt} &=& f_x(s) - g_x(x), \\
\label{S13}
\frac{dy}{dt} &=& f_y(x) - g_y(y),
\end{eqnarray}

\noindent where we have used $n_s \equiv s$, $n_x \equiv x$, and $n_y \equiv y$.
On the other hand, collecting terms of $\mathcal{O}(\Omega^{0})$ from both sides of Eq.~(\ref{S9}) results in linear Fokker-Planck equation (FPE) for the fluctuations $\epsilon_i$,
\begin{eqnarray}
\label{S14}
\frac{\partial \widetilde{P}(\pmb{\epsilon};t)}{\partial t} & = &
-\sum_{i,j=s,x,y} J_{ij} \frac{\partial}{\partial \epsilon_i} \epsilon_j \widetilde{P}(\pmb{\epsilon};t)
\nonumber \\
&& + \frac{1}{2} \sum_{i,j=s,x,y} D_{ij} 
\frac{\partial^2}{\partial \epsilon_i \partial \epsilon_j} \widetilde{P}(\pmb{\epsilon};t),
\end{eqnarray}

\noindent where, 
\begin{eqnarray}
J_{ij} &=& U_{ij}-V_{ij}, \nonumber \\
U_{ij} &=& \frac{\partial f_i(\pmb{c})}{\partial c_j} ~~~~~~ \forall i,j, \nonumber \\
V_{ij} &=& \begin{cases}
         \frac{\partial g_i(c_i)}{\partial c_i} & i=j, \\
         0 & i \neq j,
         \end{cases} \nonumber \\
D_{ij} &=& \begin{cases}
           f_i(\pmb{c}) + g_i(c_i) & i=j, \\
           0 & i \neq j.
           \end{cases}
\label{S15}
\end{eqnarray}

\noindent The average value, $\langle \epsilon_k \rangle$ at steady state can be obtained by multiplying both sides of the FPE (Eq.~(\ref{S14})) with $\epsilon_k$ and integrating over all $\pmb{\epsilon}$ \cite{Kampen2007}, one obtains at steady state,
\begin{eqnarray}
\frac{\partial \langle \epsilon_k\rangle}{\partial t} &=& 
\sum_{i=s,x,y} \langle J_{ki}\rangle \langle \epsilon_i\rangle, 
\nonumber \\
0 &=& \sum_{i=s,x,y} \langle J_{ki}\rangle \langle \epsilon_i\rangle,
\label{S16}
\end{eqnarray}

\noindent which leads to $\langle \epsilon_k\rangle =0$. Again multiplying both sides of the FPE with $\epsilon_k \epsilon_l$ and integrating over all $\pmb{\epsilon}$ \cite{Kampen2007}, at steady state we obtain,
\begin{equation}
\label{S17}
\frac{\partial \langle \epsilon_k \epsilon_l\rangle}{\partial t} =
\sum_{i=s,x,y} \langle J_{ki}\rangle \langle \epsilon_i \epsilon_l\rangle
+ \sum_{j=s,x,y} \langle J_{lj}\rangle \langle \epsilon_k \epsilon_j\rangle
+ \langle D_{kl}\rangle.
\end{equation}

\noindent The elements of covariance matrix for $\pmb{\epsilon}$ can be written as $\xi_{kl}=\langle\langle \epsilon_k \epsilon_l\rangle\rangle=\langle \epsilon_k \epsilon_l\rangle-\langle \epsilon_k\rangle \langle \epsilon_l\rangle=\langle \epsilon_k \epsilon_l\rangle$ as $\langle \pmb{\epsilon}\rangle =0$. Thus Eq.~(\ref{S17}) can be recast in the matrix form \cite{Elf2003},
\begin{equation}
\label{S18}
\frac{\partial \pmb{\xi}}{\partial t} =
\pmb{J} \pmb{\xi} + \pmb{\xi} \pmb{J}^T + \pmb{D},
\end{equation}

\noindent where, $\pmb{J}=\{ \langle J_{ij}\rangle \}$ and $\pmb{D}=\{ \langle D_{ij}\rangle \}$ are Jacobian matrix and diffusion matrix, respectively. $\pmb{J}^T$ refers to the transpose of the Jacobian matrix. At steady state, $\partial \pmb{\xi} / \partial t=0$, which leads to the Lyapunov equation for $\pmb{\xi}$,
\begin{equation}
\label{S19}
\pmb{J} \pmb{\xi} + \pmb{\xi} \pmb{J}^T + \pmb{D} = 0.
\end{equation}

\noindent Using steady state condition in Eq.~(\ref{S3}), the covariance in terms of copy number becomes,
\begin{eqnarray}
\sigma_{ij}^2 &=& \langle n_i n_j\rangle - \langle n_i\rangle \langle n_j\rangle,
\nonumber \\
&=& \Omega \langle \epsilon_i \epsilon_j\rangle,
\label{S20}
\end{eqnarray}

\noindent where we have used $\langle  \epsilon_i c_j\rangle=\langle  \epsilon_i \rangle \langle  c_j \rangle$ and $\langle c_i c_j\rangle=\langle c_i\rangle \langle c_j\rangle$. Using Eq.~(\ref{S20}) in Eq.~(\ref{S19}) we have,
\begin{equation}
\label{S21}
\pmb{J} \pmb{\sigma} + \pmb{\sigma} \pmb{J}^T + \Omega \pmb{D} = 0.
\end{equation}

\noindent Here, $\pmb{\sigma}$ is the covariance matrix with elements $\sigma_{ij}^2$. The Jacobian and diffusion matrices are expressed in concentrations $c_i$ in Eq.~(\ref{S15}). Using the definition $c_i := \lim\limits_{\Omega \to \infty} (n_i/\Omega)$ we transform the Jacobian matrix in terms of copy number $n_i$ and obtain $\pmb{J}=\{ \langle J_{ij}\rangle \}=\{ \langle U_{ij}\rangle - \langle V_{ij}\rangle \}$, where
\begin{eqnarray}
U_{ij} &=& \frac{\partial f_i(\pmb{N})}{\partial n_j} = f_{i,j}^\prime (\pmb{N}) 
~~~~~~ \forall~ i,j, \nonumber \\
\langle U_{ij}\rangle &=& f_{i,j}^\prime (\langle \pmb{N}\rangle), \nonumber \\
V_{ij} &=& \begin{cases}
         \frac{\partial g_i(n_i)}{\partial n_i} = g_{i,i}^\prime (n_i) & i=j \\
         0 & i \neq j
         \end{cases}, \nonumber \\
\langle V_{ij}\rangle &=& \begin{cases}
         g_{i,i}^\prime (\langle n_i\rangle) & i=j \\
         0 & i \neq j
         \end{cases}.
\label{S22}
\end{eqnarray}

\noindent The third term in Eq.~(\ref{S21}) can be written as $\pmb{\Xi} = \Omega \pmb{D}$, where $\pmb{\Xi}$ is the diffusion matrix at steady state in terms of the copy number $n_i$. The explicit expression of the diagonal elements of $\pmb{\Xi}$ (when $i=j$) is derived using Eq.~(\ref{S15}),
\begin{eqnarray}
\Xi_{ij} &=& \Omega \langle D_{ij}\rangle, \nonumber \\
&=& \Omega \langle f_i(\pmb{\phi}) \rangle + 
\Omega \langle g_i(\phi_i) \rangle, \nonumber \\
&=& f_i(\langle \pmb{N}\rangle) + g_i(\langle n_i\rangle).
\label{S23}
\end{eqnarray}

\noindent For $i \neq j$, $\Xi_{ij}=0$. We now rewrite the Lyapunov equation in terms of copy number as,
\begin{equation}
\label{S24}
\pmb{J} \pmb{\sigma} + \pmb{\sigma} \pmb{J}^T + \pmb{\Xi} = 0.
\end{equation}

\section{Solution of Lyapunov equation}

Using the explicit functional forms of $f$-s and $g$-s (see Fig.~1a) in Eq.~(\ref{S22}), we have the Jacobian matrix for the TSC at steady state,
\begin{eqnarray*}
\mathbf{J} = \left (
\begin{array}{ccc}
- g^{\prime}_{s,s} (\langle s \rangle) & 
0 & 
0 \\
f^{\prime}_{x,s} (\langle s \rangle) &
- g^{\prime}_{x,x} (\langle x \rangle) &
0 \\
0 &
f^{\prime}_{y,x} (\langle x \rangle) &
- g^{\prime}_{y,y} (\langle y \rangle)
\end{array}
\right ).
\end{eqnarray*}

\noindent Here, $g_{s,s}^{\prime} (\langle s \rangle)$ stands for the differentiation of $g_s (s)$ with respect to $s$ and evaluated at $s = \langle s \rangle$, and so on. In the following, we write $g_{s,s}^{\prime} (\langle s \rangle)$, $g_{x,x}^{\prime} (\langle x \rangle)$, etc as $g_{s,s}^{\prime}$, $g_{x,x}^{\prime}$, etc, respectively. Similarly, using Eq.~(\ref{S23}), the diffusion matrix can be written as,
\begin{eqnarray*}
\mathbf{\Xi} = \left (
\begin{array}{ccc}
2 g_s (\langle s \rangle) & 
0 & 
0 \\
0 &
2 g_x (\langle x \rangle) &
0 \\
0 &
0 &
2 g_y (\langle y \rangle)
\end{array}
\right ).
\end{eqnarray*}

\noindent While writing the diffusion matrix, we take into account Eqs.~(\ref{S11}-\ref{S13}) at steady state which yield $f_s=g_s(\langle s\rangle)$, $f_x(\langle s\rangle)=g_x(\langle x\rangle)$, and $f_y(\langle x\rangle)=g_y(\langle y\rangle)$.
Using the expressions of $\mathbf{J}$ and $\mathbf{\Xi}$ in the Lyapunov equation~(\ref{S24}), we have the following analytical expressions of variance and covariance for the system components at steady state,
\begin{eqnarray}
\label{S25}
\sigma_s^2 &=& \langle s\rangle, \\
\label{S26}
\sigma_x^2 &=& \langle x\rangle +
\frac{f_{x,s}^{\prime^2}}{\beta_x (\beta_s+\beta_x)} \langle s\rangle, \\
\label{S27}
\sigma_y^2 &=& \langle y\rangle +
\frac{f_{y,x}^{\prime^2}}{\beta_y (\beta_x+\beta_y)} \langle x\rangle \nonumber \\
&& +
\frac{(\beta_s+\beta_x+\beta_y) f_{y,x}^{\prime^2}}{\beta_y (\beta_s+\beta_y)(\beta_x+\beta_y)} \times
\frac{f_{x,s}^{\prime^2}}{\beta_x (\beta_s+\beta_x)} \langle s\rangle, \\
\label{S28}
\sigma_{sx}^2 &=& \frac{f_{x,s}^{\prime}}{\beta_s+\beta_x} \langle s\rangle, \\
\label{S29}
\sigma_{sy}^2 &=& \frac{f_{x,s}^{\prime} f_{y,x}^{\prime}}
{(\beta_s+\beta_x)(\beta_s+\beta_y)} \langle s\rangle, \\
\label{S30}
\sigma_{xy}^2 &=& \frac{f_{y,x}^{\prime}}{\beta_x+\beta_y} \langle x\rangle \nonumber \\
&& +
\frac{(\beta_s+\beta_x+\beta_y) f_{x,s}^{\prime^2} f_{y,x}^{\prime} }
{\beta_x(\beta_s+\beta_x)(\beta_s+\beta_y)(\beta_x+\beta_y)} \langle s\rangle.
\end{eqnarray}

\section{Noise decomposition}

The noise associated with each component (S, X, and Y) is measured by the coefficient of variation (CV). The square of CV for $i$-th component ($i \in \{s,x,y\}$) is defined as $\eta_i^2 :=\sigma_i^2/\langle i\rangle^2$. In the rest of the calculation, we use a square of CV, i.e., CV$^2$, as a metric to quantify the noise. The explicit expressions of noise associated with each component thus become,
\begin{eqnarray}
\label{S31}
\eta_s^2 &=& \underbrace{\frac{1}{\langle s\rangle}}_{\eta_{int,s}^2}, 
\\
\label{S32}
\eta_x^2 &=& \underbrace{\frac{1}{\langle x\rangle}}_{\eta_{int,x}^2} 
+
\overbrace{\underbrace{\frac{f_{x,s}^{\prime^2} \langle s\rangle^2}
{\beta_x (\beta_s+\beta_x) \langle x\rangle^2}}_{\phi_x^{int,s}} 
\times
\underbrace{\frac{1}{\langle s\rangle}}_{\eta_{int,s}^2}}^{\eta_{ext,x}^2},
\end{eqnarray}
\begin{widetext}
\begin{eqnarray}
\label{S33}
\eta_y^2 &=& \underbrace{\frac{1}{\langle y\rangle}}_{\eta_{int,y}^2} 
+
\overbrace{
\overbrace{\underbrace{\frac{f_{y,x}^{\prime^2} \langle x\rangle^2}
{\beta_y (\beta_x+\beta_y) \langle y\rangle^2}}_{\phi_y^{int,x}} \times
\underbrace{\frac{1}{\langle x\rangle}}_{\eta_{int,x}^2}}^{\eta_{ext,y1}^2} 
+
\overbrace{
\underbrace{\frac{f_{y,x}^{\prime^2} (\beta_s+\beta_x+\beta_y) \langle x\rangle^2}
{\beta_y (\beta_s+\beta_y)(\beta_x+\beta_y) \langle y\rangle^2}}_{\phi_y^{ext,x}}
\times
\underbrace{\underbrace{\frac{f_{x,s}^{\prime^2} \langle s\rangle^2}
{\beta_x (\beta_s+\beta_x) \langle x\rangle^2}}_{\phi_x^{int,s}}
\times
\underbrace{\frac{1}{\langle s\rangle}}_{\eta_{int,s}^2}}_{\eta_{ext,x}^2}
}^{\eta_{ext,y2}^2}
}^{\eta_{ext,y}^2},
\end{eqnarray}
\end{widetext}

\noindent where, $\eta_{int,i}^2$ and $\eta_{ext,i}^2$ stand for intrinsic and extrinsic noise of the component $i$, respectively. In Eqs.~(\ref{S32}-\ref{S33}), the extrinsic noises are further decomposed into more specific terms. The extrinsic noise of X is expressed as $\eta_{ext,x}^2=\phi_x^{int,s} \eta_{int,s}^2$. Similarly, the extrinsic noise of Y is $\eta_{ext,y}^2=\eta_{ext,y1}^2+\eta_{ext,y2}^2$, where, $\eta_{ext,y1}^2=\phi_y^{int,x} \eta_{int,x}^2$ and $\eta_{ext,y2}^2=\phi_y^{ext,x} \eta_{ext,x}^2=\phi_y^{ext,x} \phi_x^{int,s} \eta_{int,s}^2$. The detailed extrinsic noise components offer significant insights into the mechanism of noise propagation. Using the expressions of $f$ and $g$ given in Fig.~1a, we write the explicit analytical forms of $\phi$-s
\begin{eqnarray}
\label{S34}
\phi_x^{int,s} &=& \underbrace{
\frac{\beta_x}{\beta_s+\beta_x}}_{\tau_{x,s}}
\left(
\frac{K_{sx}}{K_{sx}+\langle s\rangle} \right)^2,
\\
\label{S35}
\phi_y^{int,x} &=& \underbrace{
\frac{\beta_y}{\beta_x+\beta_y}}_{\tau_{y,x}}
\left(
\frac{K_{xy}}{K_{xy}+\langle x\rangle} \right)^2,
\\
\label{S36}
\phi_y^{ext,x} &=& \underbrace{
\frac{\beta_s+\beta_x+\beta_y}{\beta_s+\beta_y}}_{\tau_{sy,x}^{-1}}
\frac{\beta_y}{\beta_x+\beta_y}
\left(
\frac{K_{xy}}{K_{xy}+\langle x\rangle} \right)^2,
\end{eqnarray}

\noindent where we have used the notation $\tau$ to represent the scaled time scale.

\section{Analytical expressions of channel capacity}

Under Gaussian channel approximation, the mutual information $I(i;j)$ provides the measure of channel capacity $\mathbb{C}_{ij}$ \cite{Cover1991}, which can be written in terms of signal-to-noise ratio (SNR) as \cite{Tostevin2010}, $\mathbb{C}_{ij}:=I(i;j)=(1/2)\log_2(1+\mathbb{S}_{ij})$, where $\mathbb{S}_{ij}=\eta_{ij}^4/(\eta_i^2 \eta_{j|i}^2)$ stands for the SNR. In the expression of SNR, we use $\eta_{ij}^4=\sigma_{ij}^4/(\langle i\rangle^2 \langle j\rangle^2)$, $\eta_i^2=\sigma_i^2/\langle i\rangle^2$, and $\eta_{j|i}^2=\sigma_{j|i}^2/\langle j\rangle^2$, where $\sigma_{j|i}^2=\sigma_j^2-(\sigma_{ij}^4/\sigma_i^2)$. The conditional relation can be written as $\eta_{j|i}^2=\eta_j^2-(\eta_{ij}^4/\eta_i^2)$. Using these definitions together with Eqs.~(\ref{S28}-\ref{S36}), we have
\begin{eqnarray}
\label{S37}
\eta_{sx}^4 &=& \tau_{x,s} \phi_x^{int,s} \eta_{int,s}^4,
\\
\label{S38}
\eta_{sy}^4 &=& \tau_{x,s} \tau_{y,s} \tau_{xy,s} \phi_x^{int,s} \phi_y^{ext,x} \eta_{int,s}^4,
\\
\label{S39}
\eta_{xy}^4 &=& \tau_{y,x} \phi_x^{int,s} \phi_y^{int,x} \eta_{int,x}^4
+ \tau_{y,s} \tau_{xy,s}^{-1} (\phi_x^{int,s})^2 \phi_y^{ext,x} \eta_{int,s}^4 \nonumber \\
&& + 2\tau_{y,x} \phi_x^{int,s} \phi_y^{ext,x} \eta_{int,s}^2 \eta_{int,x}^2,
\end{eqnarray}

\noindent where, $\tau_{i,j}=\beta_i/(\beta_i+\beta_j)$ and $\tau_{ij,k}=(\beta_i+\beta_j)/(\beta_i+\beta_j+\beta_k)$. The explicit expressions of SNR for the TSC motif thus become
\begin{widetext}
\begin{eqnarray}
\label{S40}
\mathbb{S}_{sx} &=& \frac{\tau_{x,s} \phi_x^{int,s} \eta_{int,s}^2}
{\eta_{int,x}^2+(1-\tau_{x,s}) \phi_x^{int,s} \eta_{int,s}^2},
\\
\label{S41}
\mathbb{S}_{sy} &=& \frac{\tau_{x,s} \tau_{y,s} \tau_{xy,s} \phi_x^{int,s} \phi_y^{ext,x} \eta_{int,s}^2}
{\eta_{int,y}^2 + \phi_y^{int,x} \eta_{int,x}^2 + (1-\tau_{x,s} \tau_{y,s} \tau_{xy,s}) \phi_x^{int,s} \phi_y^{ext,x} \eta_{int,s}^2},
\\
\label{S42}
\mathbb{S}_{xy} &=& \frac{
\tau_{y,x} \phi_y^{int,x} \eta_{int,x}^4
+ \tau_{y,s} \tau_{xy,s}^{-1} (\phi_x^{int,s})^2 \phi_y^{ext,x} \eta_{int,s}^4
+ 2\tau_{y,x} \phi_x^{int,s} \phi_y^{ext,x} \eta_{int,s}^2 \eta_{int,x}^2
}{\left[\splitdfrac{\eta_{int,x}^2 \eta_{int,y}^2 + 
\phi_x^{int,s} \eta_{int,s}^2 (\eta_{int,y}^2 + \phi_y^{int,x} \eta_{int,x}^2)
+ (1-\tau_{y,x}) \phi_y^{int,x} \eta_{int,x}^4}
{+ (1-\tau_{y,s} \tau_{xy,s}^{-1}) (\phi_x^{int,s})^2 \phi_y^{ext,x} \eta_{int,s}^4
+ (1-2\tau_{y,x}) \phi_x^{int,s} \phi_y^{ext,x} \eta_{int,s}^2 \eta_{int,x}^2
} \right]}.
\end{eqnarray}
\end{widetext}

In the present study, we use separation of degradation time scales $\beta_s < \beta_x < \beta_y$. The inequality in $\beta$-s results in $\tau_{x,s} \approx 1$, $\tau_{y,s} \approx 1$, $\tau_{xy,s} \approx 1$, and $\tau_{y,x} \approx 1$, for which we approximate $1-\tau_{x,s} \approx 0$, $1-\tau_{x,s} \tau_{y,s} \tau_{xy,s} \approx 0$, $1-\tau_{y,s} \tau_{xy,s}^{-1} \approx 0$, and $1-2\tau_{y,x} \approx -1$. Using these approximations, the expressions of SNR yield,
\begin{eqnarray}
\label{S43}
\mathbb{S}_{sx} &=& 
\frac{\phi_x^{int,s} \eta_{int,s}^2}
{\eta_{int,x}^2},
\\
\label{S44}
\mathbb{S}_{sy} &=& 
\frac{\phi_x^{int,s} \phi_y^{ext,x} \eta_{int,s}^2}
{\eta_{int,y}^2 + \phi_y^{int,x} \eta_{int,x}^2},
\\
\label{S45}
\mathbb{S}_{xy} &=& \frac{A}{B},
\end{eqnarray}

\noindent where, 
\begin{eqnarray*}
A &=& \phi_y^{int,x}\eta_{int,x}^4 + (\phi_x^{int,s} \eta_{int,s}^2 + 2 \eta_{int,x}^2) \phi_x^{int,s} \phi_y^{ext,x} \eta_{int,s}^2, \\
B &=& \eta_{int,x}^2 \eta_{int,y}^2 + (\eta_{int,y}^2+\phi_y^{int,x}\eta_{int,x}^2) \phi_x^{int,s} \eta_{int,s}^2 \nonumber \\
&& - \phi_x^{int,s} \phi_y^{ext,x} \eta_{int,s}^2 \eta_{int,x}^2.
\end{eqnarray*}

Using Eqs.~(\ref{S43}-\ref{S45}), the channel capacities are written as,
\begin{eqnarray}
\mathbb{C}_{sx} &=& \frac{1}{2} \log_2 \left[
1 + \frac{\phi_x^{int,s} \eta_{int,s}^2}{\eta_{int,x}^2}
\right], \label{S46} \\
\mathbb{C}_{sy} &=& \frac{1}{2} \log_2 \left[
1 + \frac{\phi_x^{int,s} \phi_y^{ext,x} \eta_{int,s}^2}
{\eta_{int,y}^2 + \phi_y^{int,x} \eta_{int,x}^2}
\right], \label{S47} \\
\mathbb{C}_{xy} &=& \frac{1}{2}
\log_2 \left[ 1+
\frac{A}{B} \right].
\label{eqnS48}
\end{eqnarray}


%

\end{document}